\documentclass[twocolumn,prl]{revtex4-1}
\usepackage{graphicx}
\usepackage{amsmath}
\usepackage{color}
\usepackage{subfigure}
\usepackage{latexsym}
\def\be{\begin{equation}}
\def\ee{\end{equation}}
\def\bea{\begin{eqnarray}}
\def\eea{\end{eqnarray}}
\def\a{\alpha}
\def\b{\beta}
\def\p{\partial}
\def\besub{\begin{subequations}}
\def\eesub{\end{subequations}}
\def\n{{\bf n}}

\def\nn{\nonumber}
\def\l{\left(}
\def\r{\right)}

\def\th{\theta}
\def\rar{\rightarrow}
\def\f{\frac}

\def\bwd{\begin{widetext}}
\def\ewd{\end{widetext}}

\begin{document}

\title{Patterning of polar active filaments on a tense cylindrical membrane}
\author{Pragya Srivastava$^1$, Roie Shlomovitz$^2$, Nir S.Gov$^3$ and Madan Rao$^{1,4}$}
\affiliation{$^1$Raman Research Institute, Bangalore 560080, India\\
$^2$Department of Physics and Astronomy, University of California, Los Angeles, 
USA \\
$^3$Chemical Physics, Weizmann Institute of Science, Rehovot, Israel \\
$^4$National Centre for Biological Sciences (TIFR), Bangalore 560065, India\\ }
\date{\today}
\begin{abstract}
We study the dynamics and patterning of polar contractile filaments on the surface of a cylindrical cell using active hydrodynamic equations that incorporate couplings between curvature and filament orientation. 
Cables and rings spontaneously emerge as steady state configurations on the cylinder, and can be stationary or moving, helical or segments moving along helical trajectories. 
Contractility induces coalescence of proximal rings. We observe phase transitions in the steady state patterns upon changing cell diameter and make several testable predictions.  Our results are relevant to 
the dynamics and patterning of a variety of active biopolymers in cylindrical cells.
\end{abstract} 
\maketitle

Cytokinesis, a common mechanism by which cells divide, involves the regulated assembly and constriction of a contractile actomyosin ring. 
A well studied model system of the dynamics of assembly of the contractile ring is {\it fission yeast} \cite{pollardreview}. The contractile ring appears to assemble on the inner cell surface via heterogeneous nucleation of {\it nodes}  and growth in the form of {\it actin cables} \cite{pollardreview}.
Recent experiments on cylindrical fission yeast and its mutants have shown that the stability and dynamics of the contractile ring is greatly influenced by cell geometry \cite{JCellSc}.
In this paper, we account for these different phenotypes exhibited by the actomyosin filaments using a single model involving the
interplay between geometry and active contractile mechanics. Indeed this interplay between geometry and 
cytoskeletal mechanics has been well appreciated in the context of mitotic division in epithelial cells \cite{Odde}.

\begin{figure}[ht]
  \begin{center}
    \subfigure{\includegraphics[height=3.5in]{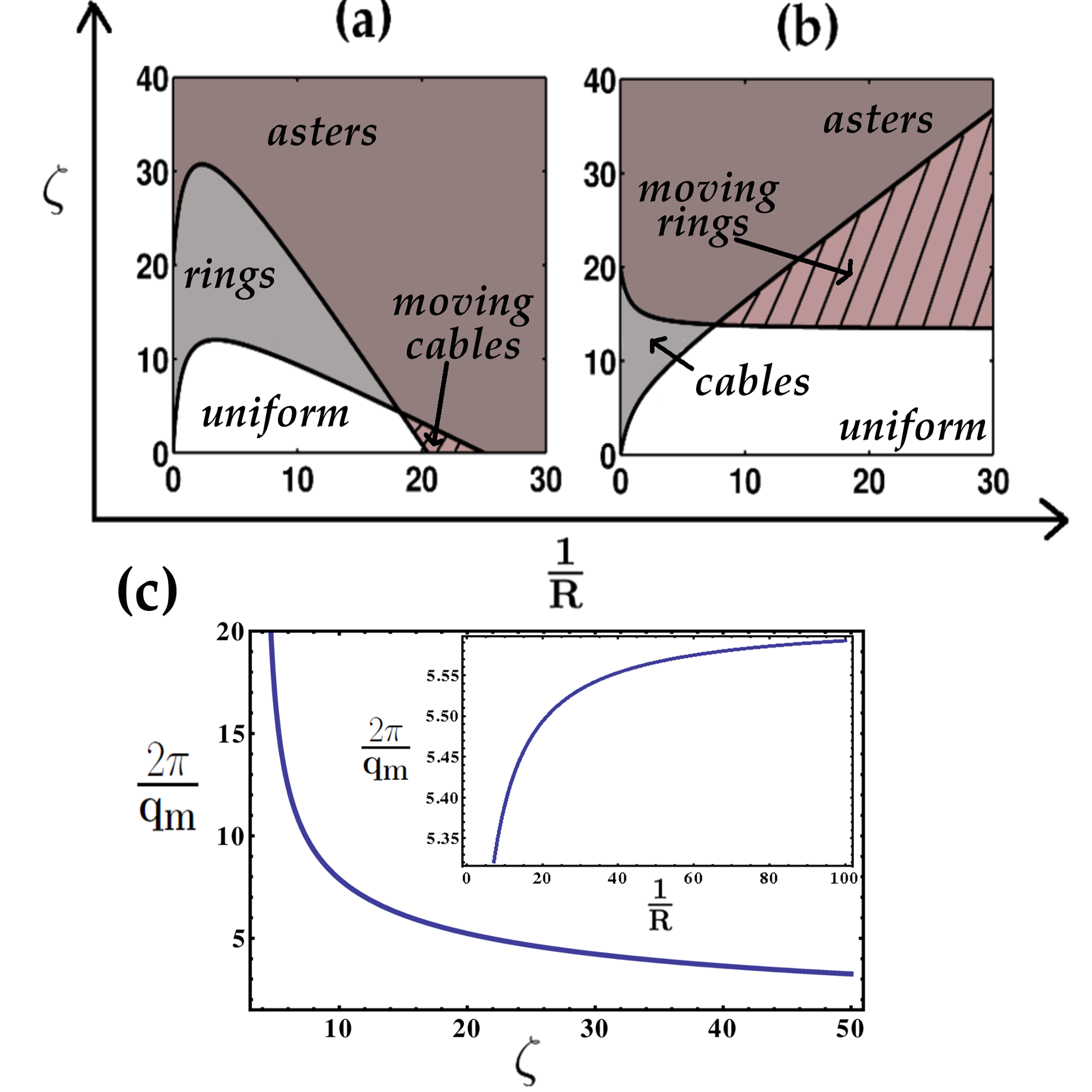}}
    \end{center}
\caption{(color online)  
Phase diagram in $\zeta - R$ at high filament concentration and net mean orientation, obtained from linear stability analysis, for (a) $\a_\th>0$ and $\a_\th>\a_z$ and (b) $\a_z>0$ and $\a_z>\a_\th$. The steady state patterns corresponding to these phases are depicted in Fig.\,2. Uniform phase refers to homogeneous, oriented phase along (a) $\theta$
and (b) $z$.
 (c) Scaling of the inverse of the fastest growing wave vector $q_{m}$ with  $\zeta$ and $R$ (inset). All quantities are in scaled units
 \cite{Supp}.
}  
\label{phdia}
\end{figure}

We present an extensive analysis of the steady state patterns of active polar filaments  (e.g., actomyosin) on the (inner) surface of  a nondeformable cylindrical cell, using the equations of \emph{active hydrodynamics} \cite{activereview} generalized to include the coupling of geometry to filament orientation \cite{note}.  Such nondeformable cylindrical cell surfaces are realized in rod-shaped bacteria and fission yeast cells, due to their strong coupling 
to a rigid cell wall \cite{pollardreview}.
While our study complements the approach based on microscopic agent-based simulations \cite{shaugh,vavylonis}, it has significant
points of difference, notably the influence of cell geometry and size. Our main results are :
(a)  Rings and cables appear as generic steady state patterns, either from a spinodal instability of the homogenous configuration 
or from a nucleation and growth of an actin node. Rings/cables may be either stationary or mobile, and the 
 ring/cable width is set by a Pecl\'et length $L_p$, the ratio of the filament diffusion to a curvature-dependent active filament advection. 
(b) Phase transitions between the different steady state configurations can be achieved by changing cell diameter (Fig.\,\ref{phdia}). (c) Due to the intrinsic anisotropy in curvature of the cylindrical cell, the active filaments 
can exhibit helical shapes or tilted segments which move along helical trajectories. 
(d) Proximal rings  merge to form one ring, the
dynamics of ring merger is controlled by actomyosin contractility.

We describe the dynamics of active polar filaments in a 2-dimensional cylindrical surface  ${\cal S}$, in terms of a local concentration $c({\bf r},t)$ and polarization $\n\l{\bf r},t\r
=\l n_\th, n_z \r 
$, representing the polar orientation (equivalently, the velocity of the filaments relative to the medium). Assuming that momentum gets dissipated predominantly by friction at the cortex, the hydrodynamic velocity is obtained from local force balance,
$-\Gamma {\bf v} = \nabla \cdot \sigma $ where $\sigma = - W c \n \n$ is active contractile stress with $W<0$ \cite{hatwalne}. Conservation of total number of actin filaments in the cortex
 implies,
$\partial_t c = - \nabla \cdot {\bf J} $,
where, $\nabla \equiv \l R^{-1}\p_{\th},  \p_z \r$ is the differential operator on the surface of cylinder of radius $R$.
The filament current $ J_i =   v_0  c n_i + \Lambda_{ijkl} \kappa_{jk} c n_l - D \nabla_i c $, where the first term is 
an active advection (e.g., motor-driven or (de)polymerization) with speed $v_0$, the second is an anisotropic advection  due to coupling between $\n$ and the curvature tensor $\kappa_{ij}$, and the third is a diffusive current with coefficient $D$. 
We will implicitly assume that the filaments are short compared to the cell size.

The dynamics of the polarization $\n$ is obtained by generalizing the Toner-Tu equations \cite{tonertu} to include symmetry allowed couplings to cell geometry,
\bea
\nn
\f{\p \n}{\p t} + \lambda \l \n \cdot \nabla \r \n  & = &  K_1 \nabla^2 \n + K_2 \nabla \l\nabla \cdot \n \r  \\ \nn
& & + \zeta \nabla c + \l \a- \beta \vert n \vert^2\r {\bf n}+ \mbox{\boldmath$\gamma\, \kappa$}  \, \n \\
\label{neq}
\eea
where the laplacian is defined on the surface of the cylinder, and the
last term implies tensor contraction. 

The $4^{th}$ rank tensor parameters $\Lambda^{ij}_{kl}$ and $\gamma^{ij}_{kl}$ are the most general phenomenological coupling between curvature $\kappa_{ij}$ and polar order $\n$ to lowest order.  While the coupling parameter $\gamma^{ij}_{kl}$ can be present even in an equilibrium membrane with a tilt field, the parameter $\Lambda^{ij}_{kl}$ is purely active in origin. Using symmetry considerations and the form of the curvature tensor  $\kappa = \tiny{\left(\begin{array}{cc}1/R & 0 \\0 & 0\end{array}\right)}$ for a cylinder, we find there are two independent non-zero contributions from $\Lambda^{ij}_{kl}$ and $\gamma^{ij}_{kl}$ each,  denoted as $\Lambda_{i}$ and $\gamma_{i}$ with $i=\theta,z$, henceforth. Note that $\Lambda_{i}$ trivially renormalize the advection $v_0$ in the $c$ equation, 
giving rise to an anisotropic advection $v_{i}=v_0 + \Lambda_{i}/R$. Similarly, $\gamma_{i}$ renormalize the coefficient of the linear term $\alpha$ in  Eq. \ref{neq} to $\alpha_{i}=\alpha + \gamma_{i}/R- \delta_{i \theta}K_1/R^2$ where $\delta_{ij}$ is the Kronecker delta.
In what follows, we let $\gamma_{i}$ to have positive and negative values while we restrict  $K_{1,2}$ to the positive domain.

The parameter $\alpha$ is a linear function of the mean filament concentration $c_0$ and governs the transition from orientationally disordered to polar ordered phase, while $\zeta>0$ describes the tendency of contractile filaments to reorient towards the gradient of concentration \cite{kripa}. Due to activity of the filaments they move relative to the 
solvent  in the direction of ${\bf n}$, thus reorientation is accompanied by movement of filaments towards each other and can, for instance, be enhanced by increasing motor activity.
We explore the steady state patterns of active filaments on a cylindrical cell as a function of (i) $c_0$, (ii) $\zeta$ and (iii) cell size $R$. Values of parameters used correspond to those in fission yeast ({\it S. Pombe}), and are tabulated in \cite{Supp}.



We first study the spinodal instability of the homogeneous configuration with concentration $c_0$, which is stable when $\zeta=0$. When $\alpha_{i}(c_0,R) < 0$, i.e., when the filament concentration is low enough,
the stable homogeneous phase has no orientational order, $\langle \n \rangle =0$.  We ask whether the homogeneous-isotropic phase  $(c=c_0, \n={\bf 0})$ continues to be stable as we increase $\zeta$. Using linear stability analysis we find that for $\zeta > \zeta_c(\{\alpha_{i}\}; \{v_{i}\})$  this phase become unstable with a maximally unstable wavevector ${\bf q}_{m}$  along either  the $\th$ or the $z$ axis \cite{Supp}, this gives rise to the phase boundaries depicted in \cite{Supp}.
The nature of the final steady state configurations characterizing the phases, however, cannot be obtained from a linear analysis. 
To ascertain this we numerically integrate \cite{Supp} the dynamical equations for $c$ and ${\bf n}$  using an implicit and alternating direction method \cite{Numrec}. The finite wavevector  instability seen in the linear analysis shows up as a density clumping along a specific  direction.
For parameter ranges which result in only one maxima in the dispersion surface, the direction of the density clumping is set by
the direction of the
 fastest growing wavevector ${\bf q}_{m}$  \cite{Supp}, thus
when ${\bf q}_{m}$ is along  $z$, we obtain {\it rings},
while when ${\bf q}_{m}$ is along  $\th$, we get {\it cables}.
For the parameter range where the dispersion surface shows two maxima (along both $q_z$ and $q_\th$), the final configurations correspond to  {\it asters/nodes} \cite{Supp}.

When the filament concentration is high enough so that  one or both of the $\alpha$'s are positive ($\alpha_{i}(c_0,R)  > 0$), the homogeneous solution $c=c_0$ has polar order, $\vert \langle \n \rangle \vert \neq0$ (Fig.\ref{phdia}). The spontaneous polarization is determined  by the largest  $\alpha$ and can be along either the azimuthal ($\theta$-axis) or the axial ($z$-axis) direction. Again we use linear stability analysis to check the stability of the uniform solution upon increasing $\zeta$.
Since the nature of instability is the same whether $\alpha_{\theta}>\alpha_{z}$ or vice versa, we discuss, without any loss of generality,  the former case where we perturb the system around  $n_0 =\sqrt{\a_\th/\b}$. The system tends to  form spatial structures  only beyond a threshold contractility $\zeta$ given below.
  As before, the orientation of the spatial structure is set by the fastest growing wavevector ${\bf q}_{m}$ of the dispersion surface : (a) when $\zeta_{c2} > \zeta > \zeta_{c1} = \f{D \l \a_\th- \a_z \r}{c_0 v_z}$,
   the instability
  is along $z$, while 
  (b) when $\zeta_{c1} > \zeta>\zeta_{c2} = \f{2 \a_\th}{c_0 v_\th}$, the instability is  along $\th$.
Fig.\ref{phdia}(c) shows the dependence of the magnitude of the fastest growing wavevector on $R$ and $\zeta$, the exact expressions are displayed in \cite{Supp}. 

Once again, linear stability analysis tells us nothing about the final steady state configuration, which we have to obtain by numerical integration of the dynamical equations. We confirm that the linear instabilities in the paramater ranges (a) and (b), give rise to periodically separated stationary rings and moving cables, respectively (Fig.\,2). However, when $\zeta > \max(\zeta_{c2}, \zeta_{c1})$, the dispersion surface has two maxima, one along $z$ and the other along $\th$ \cite{Supp}, 
 and a numerical solution of the dynamical equations, shows that the steady state configuration is an array of inward-pointing asters/nodes. The final  phase diagram in $R, \zeta$ is shown in Fig.\,\ref{phdia}(a), where the phase boundaries are obtained from linear stability analysis, while the nature of the final steady state configurations are obtained numerically. The time sequence of configurations
 starting from the homogeneous, oriented phase and leading to the formation of rings is displayed in \cite{Supp}.
 The phase diagram for the case when 
 $\a_z >0$ and $\a_z>\a_\th$ (i.e., when the spontaneous polarization is along $z$) is shown in 
 Fig.\,\ref{phdia}(b). 
 
From the numerics we find that the magnitude of ${\bf q}_{m}^{-1}$ corresponds to the periodicity $d$ of the rings,  leading  to two testable predictions (Fig.\,\ref{phdia}(c)) : (i) the distance between rings $d$ increases monotonically with $R$ and saturates to a
 constant which depends on $\zeta$  \cite{Supp} and (ii) $d$ decreases with contractility as $1/\sqrt{\zeta}$ \cite{Supp}.


\begin{figure}[ht]
 \begin{center}
  \subfigure{\includegraphics[height=1.0in]{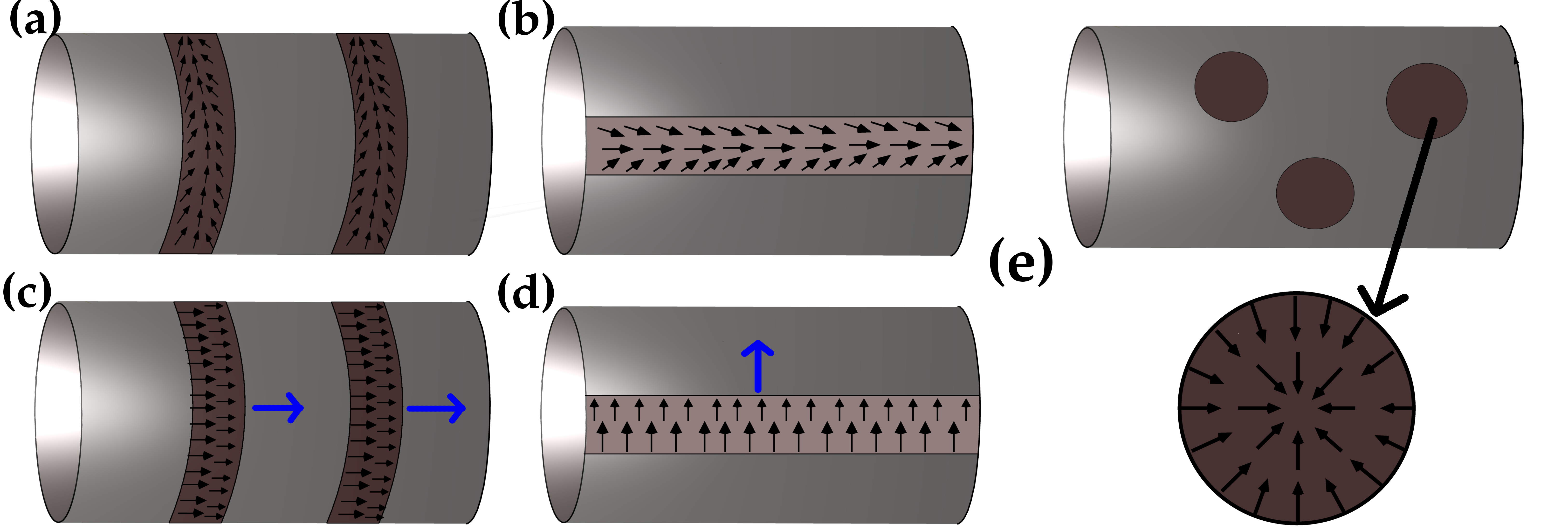}}
   \end{center}
\caption{(color online) Schematic of steady state patterns characterizing the phases in Fig.1 : (a) stationary rings (b) stationary cables, (c) moving rings, (d) moving cables (e) asters/nodes. Small arrows (black) show texture and thick arrow (blue) denotes direction of movement of the rings/cables.  } 
\label{ener}
\end{figure}

The width $w$ of the rings and cables (and aster size) is set by the ratio $D/v_{\th,z}$, the curvature-dependent active Pecl\'et length, which balances diffusion with advection. To obtain the leading contribution to the ring velocity, we use the ansatz for an axisymmetric moving ring $c\equiv c\l z-z_0(t)\r$ together with the  assumption that the other variables, such as $w$ and the polarization-profile are {\it fast}; integration over the ring area then gives $\dot{z}_0(t)= v_z(\zeta,R)$.  It is reasonable to expect $\Lambda_z >0$; this  leads to another prediction of the model, namely the velocity of the moving ring is higher for narrower cylinders and approaches $v_0$ as $1/R$, when $R$ is large. Similar arguments suggest that the velocity of moving cables is $v_\th \l\zeta, R \r$.

We have implicitly assumed that $K_{1,2} \geq0$, thus obtaining parallel filament  orientations within the ring. However the steady states also admit {\it anti-parallel} filament orientations if $K_{1,2} < 0$; one then needs to augment 
Eq.(\ref{neq}) by a symmetry allowed 4th-order spatial derivative for stability.

Thus far we have studied the spinodal instability of the uniform phase to coherent structures. In the context of fission yeast however, actin is {\it nucleated} at the cell surface by the actin nucleator {\it formin} \cite{pollardreview},
which first  forms a node, then grows into a ring or cable. We study this `nucleation and growth' driven transition as a function of $c_0$, $\zeta$ and $R$. This is easiest to study in the limit $\lambda \rar 0$, when the 
 the dynamics of ${\bf n}$ can be obtained from an ``energy'' functional (strictly lyapunov functional),
\be
E\left[c, \n\right]= \int_{{\cal S}}  \l K_1 + K_2 \r  \l \nabla \cdot \n \r^2+ K_1 \l\nabla \times \n \r^2 + \zeta c \l \nabla \cdot \n \r, 
\label{aniso_free}
\ee
together with a local constraint on the magnitude, ${\bf n}\cdot {\bf n}\equiv n_{0}^2=\alpha_{\th,z}/\beta$.
We now compare the ``energies'' of an inward-aster ($E_a$), a $\th$-segment ($E_{\th}$)  and a $z$-segment ($E_z$) of fixed area $A$ (where the segment widths are determined by the balance of the
current ${\bf J}$). We make the simplifying assumption that the concentration 
within these regions is uniform and so we are comparing configurations with the same mean concentration $c_0$. In one constant approximation, $K_1=K_2=K$,
\besub
\bea
E_{a}&=&  \pi K \ln \f{A}{\pi\xi^2 }- 4 \pi \zeta c_0 \sqrt{\f{A}{\pi}} + \epsilon_c \\ 
E_{\th,z}&=& \left[ \f{\pi^2K}{D^2} \l v_0 + \f{\Lambda_{z,\th}}{R}\r^2 -  \f{2\zeta c_0}{D} \l v_0 + \f{\Lambda_{z,\th}}{R} \r \right] A\nn\\
\label{nucleation}
\eea
\eesub
where  $\xi$ and $\epsilon_c$ are the core size and energy, respectively.  Note that $E_a$ is the same as in a planar geometry, since the gaussian curvature of a cylinder is zero \cite{gausscurv}. The ``energy'' branches and the
phase diagram are shown in Fig.(\ref{ener}).

\begin{figure}[ht]
 \begin{center}
  \subfigure{\includegraphics[height=3.25in]{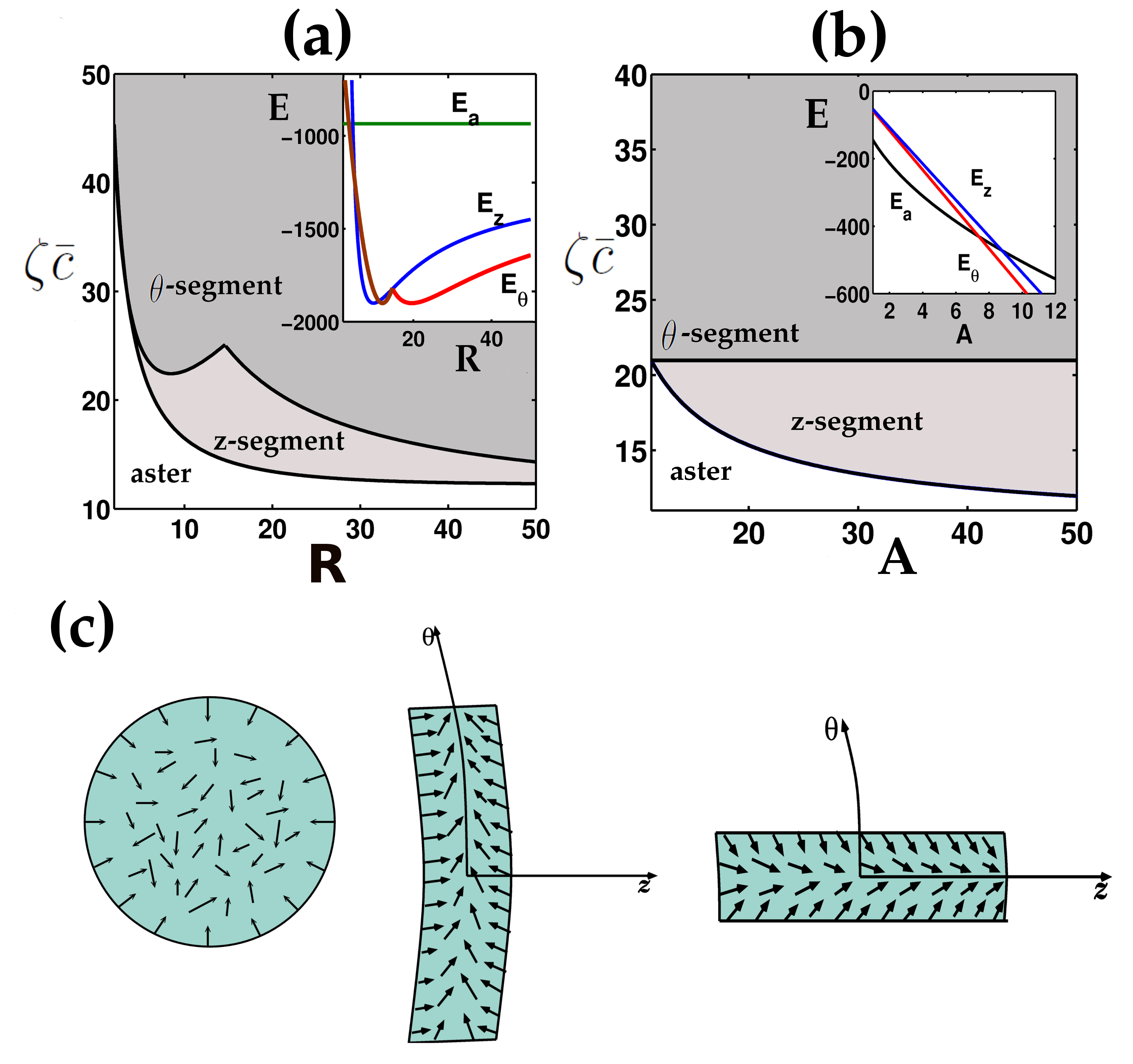}}
   \end{center}
\caption{(color online) Phase diagram in $\zeta c_0$ versus (a)  $R$ (with $A$ constant) and (b) $A$ (with $R$ constant) showing asters (nodes), $\th$-segment and $z$-segment. Insets show the
``energy'' branches (units of $K$) of the 3 configurations, Eq.(\ref{nucleation}).  Note that the form of $E_{\th}$ in Eq.(\ref{nucleation}) is valid for  a $\th$-segment which does not encircle the cylinder,  for which the
segment width is set by $D/v_z$. For smaller values of $R$ (below the kink in (a)), the $\th$-segment forms a ring, whose width is set by $A/2\pi R$. The orientation of ${\bf n}$ in these configurations are shown 
in (c).} 
\label{ener}
\end{figure}

\begin{figure}[ht]
 \begin{center}
  \subfigure{\includegraphics[height=1.0in]{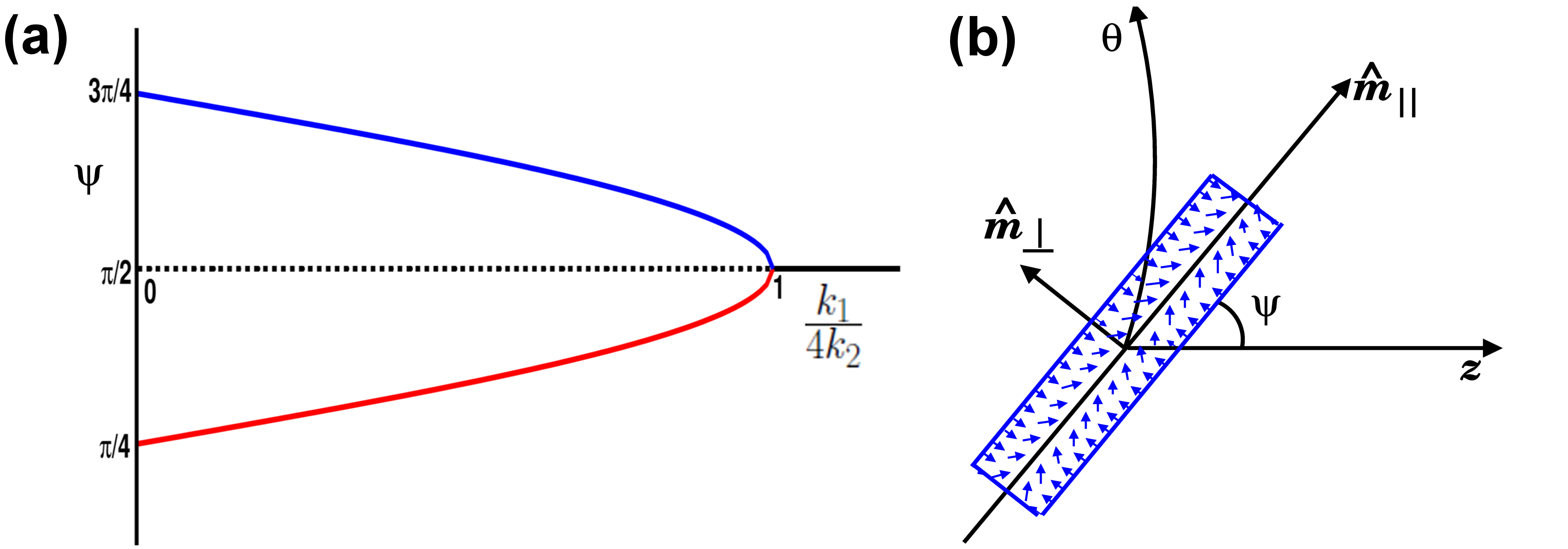}}
   \end{center}
\caption{(a) Orientation of tilt-segment $\psi$, schematically represented in (b), with (short) arrows showing $\n$ within the segment. The steady state
tilt plotted as a function of $\f{k_1}{4 k_2}$ shows a continuous transition from a $\th$-segment ($\psi=\pm \f{\pi}{2}$)  to a tilt-segment (see text) at $\f{k_1}{4 k_2}=1$. The symmetries of a cylinder
allow for $4$ stable solutions, the two shown above have a net orientation of $\n$ along $\psi$, while two more have a net orientation of $\n$ along $-\psi$.  
} 
\label{helix}
\end{figure}

Thus far, the steady state configurations correspond to rings (cables) along the $\th$ ($z$ axis) alone, no other orientation is permitted. However
in principle, the parameters accompanying the {\it nonlocal} terms in (\ref{neq}), namely $K_1,K_2$ and $\zeta$, can also have anisotropies consistent with cylindrical symmetry.
This could arise, for instance, from having ``easy-directions'' \cite{spinanisotropy} for spatial variations of ${\bf n}$ and $c$.

Consider a tilted segment at an angle $\psi$ to the cylinder axis ($\psi =\cos^{-1}({\bf {\hat m}}_{\|}\cdot {\bf {\hat z}})$, where ${\bf {\hat m}}_{\|}$ is the unit vector along the segment boundary, Fig.(\ref{helix}b)),
 having a uniform width $w(\psi)$ and length $l(\psi)$ such that the area $A=w l$ is fixed.
For simplicity, we assume that within the tilt-segment, the filament concentration $c=c_0$ is uniform. The texture within the tilt-segment is shown in Fig.(\ref{helix}b), the orientation of $\n$
 changes by $\pi$ on a length scale $w(\psi)$.
The width of the tilt-segment $w(\psi)$ is set by a balance of the net current ${\bf J}\cdot{\bf {\hat m}}_{\perp}$, where ${\bf {\hat m}}_{\perp}$ is a unit vector normal to the boundary.
In the equal constants approximation, $K_1=K_2=K$, the ``energy'' (\ref{aniso_free}) of this tilt-segment is given by, 
\be
E(\psi)=K(\psi) \f{\pi^2 A}{w(\psi)^2}  - 2  \zeta(\psi) {c_0}   \f{A}{w(\psi)} 
\ee
and the steady state orientation of the tilt-segment is  obtained by setting $\f{\p E}{\p \psi}{\big\vert}_A=0$. 
To prove our point, it suffices to look at the anisotropy of $K(\psi)$ alone,  which owing to cylindrical symmetry
can be written as $K(\psi)=\sum_m k_m \cos 2 m \psi$, of which we take only the first two modes.
The phase diagram Fig.(\ref{helix}), shows a continuous transition from a $\th$-segment with $\psi=\pm \pi/2$, to a tilt-segment
with $\psi=\cos^{-1} \sqrt{\f{1}{2}\l1 - \f{k_1}{4 k_2}\r}$, as $k_1/4k_2$ is varied. Because the net polarization of these {\it active}  tilt-segments is along $\psi$, the short tilt-segments {\it move} on helical tracks, while the longer 
tilted segments appear as moving helices (see Refs.\cite{FtsZExpt,FtsZThe} for propagating helices of FtsZ
in B.{\it subtilis}). The analysis is relevant to recent experiments \cite{mreb} on the movement of MreB filaments  
 on helical tracks, whose strong anchoring to the cell wall growth machinery provides the ``easy-directions'' on the surface of these rod-shaped bacteria.

We end with a discussion on the fate of multiple rings, both moving and stationary. Well separated rings {\it moving} along the $+z$-axis, result in periodic oscillations in time (with a time period $\tau =L/v_z$, which gets smaller for narrower cylinders), when we impose
no flux boundary conditions along $z$ or when the cylinder is capped on either end by hemispheres. On the other hand,  {\it stationary} rings, proximal to each other, merge to eventually form a single
ring whose width is the same as original rings \cite{Erikson}. Our analytical and numerical studies suggest that such ring merger occurs whenever there is a nonzero overlap of the concentration profiles of the rings;
the dynamics of merger shows an initial slow rate governed by $K$ which crosses over to a faster rate given by the contractile parameters, $\sqrt{\zeta v_z}$ (which gets larger for narrower cylinders).

In summary, we have presented a detailed analysis of the effect of curvature-orientation coupling on the patterning of active filaments in a cylindrical cell and predict a rich variety of steady state patterns which include asters (nodes), stationary and moving rings and cables.  Our work is valid in those situations where we can ignore the deformation of the cell membrane, which can itself contribute to additional
instabilities \cite{Roie}.
The coarse-grained symmetry based approach presented here is complementary to the more microscopic agent-based simulations, and has the advantage of being applicable to a variety of cellular contexts such as
 (i) actin filaments in fission yeast cells \cite{pollardreview}, (ii) reconstituted cytoskeletal elements on cylindrical liposomes \cite{Erikson}, (iii) FtsZ filaments in bacteria (and reconstituted in yeast \cite{srinivasan}) and (iv) MreB filaments in rod shaped bacteria \cite{mreb}. 
 A more detailed comparison to experiments will be taken up later.
Our work sets the stage to study the dynamics of {\it quenches} from one phase to another, which is appropriate to the study of the growth of cables or rings from the nucleation of nodes of acto-myosin filaments in fission yeast \cite{pollard,HFSP}.

We thank M. Balasubramanian and his group for discussions and collaborations. MR thanks HFSP and NSG  thanks the support of the Minerva Foundation (grant no. 710589).

\end{document}


\title{Patterning of polar active filaments on a tense cylindrical membrane\\ P.Srivastava et al. : Supplementary Material}
\begin{abstract}
We provide a table containing the values of parameters used in the analysis and a description of the numerical scheme. We also provide some details  of the linear stability analysis with analytic expressions and dispersion plots  in various parameter regimes. We have also included a time sequence of configurations starting from the unstable homogeneous phase to the formation of rings.
\end{abstract} 
\maketitle

\section{Units and Parameter values}
We have made the dynamical equations dimensionless by choosing 
appropriate units for length, time and order parameter (polarization, $n$),
\begin{enumerate}
\item Unit of length, $l = \frac{R}{5}$. Taking $R= 2 \, \mu\text{m}$ for fission yeast \cite{pollard}, this translates to
$l = 0.4\mu$m. All length scales are measured in units of $l$.

\item Unit of  time, $\tau = \frac{R}{5 v_0}$. Taking $v_0=0.4\mu ms^{-1}$ for fission yeast \citep{lord},  this translates to
$\tau= 1$\,s. All time scales are measured in units of $\tau$.

\item Unit of magnitude of polarization, $n=\sqrt{\alpha/\beta}$. The values of $\alpha$ and $\beta$ have been chosen to be large ($=10$) 
and equal so as to ensure that this magnitude is $1$ almost everywhere. 

\end{enumerate}

To obtain the values of the parameters in real units, we multiply the number used in the simulation by its appropriate dimension in
units of $\ell$, $\tau$ and $n$ (see Table below).\\


%

\begin{tabular}{|l|c|r|}
\hline 
Parameters& Physical values & Scaled values \\ \hline
$R $ &2 $\mu$m \cite{pollard} &5 $(l)$ \\ \hline
$L $ &15 $\mu$m \cite{pollard}& $37.5 (l)$ \\ \hline
$D $ &$0.1-1 \mu \text{m}^2s^{-1}$ \cite{carl} & $0.625-6.25 (l)$ \\ \hline
$v_0 $& 0.4 $\mu\text{m}s^{-1}$ \cite{lord} & 1 $(l/\tau)$\\ \hline
$K_1$ & 0.16 $\mu\text{m}^2s^{-1}$  & $1 (l^2/\tau)$\\ \hline
$K_2 $ &0 $\mu\text{m}^2s^{-1}$ &$0(l^2/\tau)$ \\ \hline
$\a $& 10 $s^{-1}$ & $10 (\tau^{-1})$ \\ \hline
$\gamma_{i} $& $-20-10 \mu\text{m}s^{-1}$ & $-50-25  (l/\tau) $ \\ \hline
$\Lambda_{i} $& $-0.8-1.2 \mu\text{m}^2s^{-1}$ & $-5-7.5 (l^2/\tau)$\\ \hline
\end{tabular}
\\

\vspace{0.1in}
Note that in the above Table, $K_1$ has been chosen to be of same order as $D$. For simplicity, we have taken  $K_2=0$. Values of $\gamma_{i}$ and $\Lambda_{i}$ are chosen such that $\f{\gamma_{i}}{R}$ and $\f{\Lambda_{i}}{R}$ are of same order as $\a$ and $v_0$. 
\\

\section{Numerical Scheme}
We numerically integrate the dynamical equations  using an implicit scheme with alternating direction method \cite{Numrec} with timestep $\Delta t= 0.01$ and gridsize $\Delta z =0.375$ and $R \Delta \theta =0.315$, in units of $\tau$ and $l$. 
\\

\section{Dispersion relations and fastest growing wave vectors}
\subsection{Low mean concentration of polar filaments}

\begin{figure*}[ht]
  \begin{center}
    \subfigure{\includegraphics[height=2.3in]{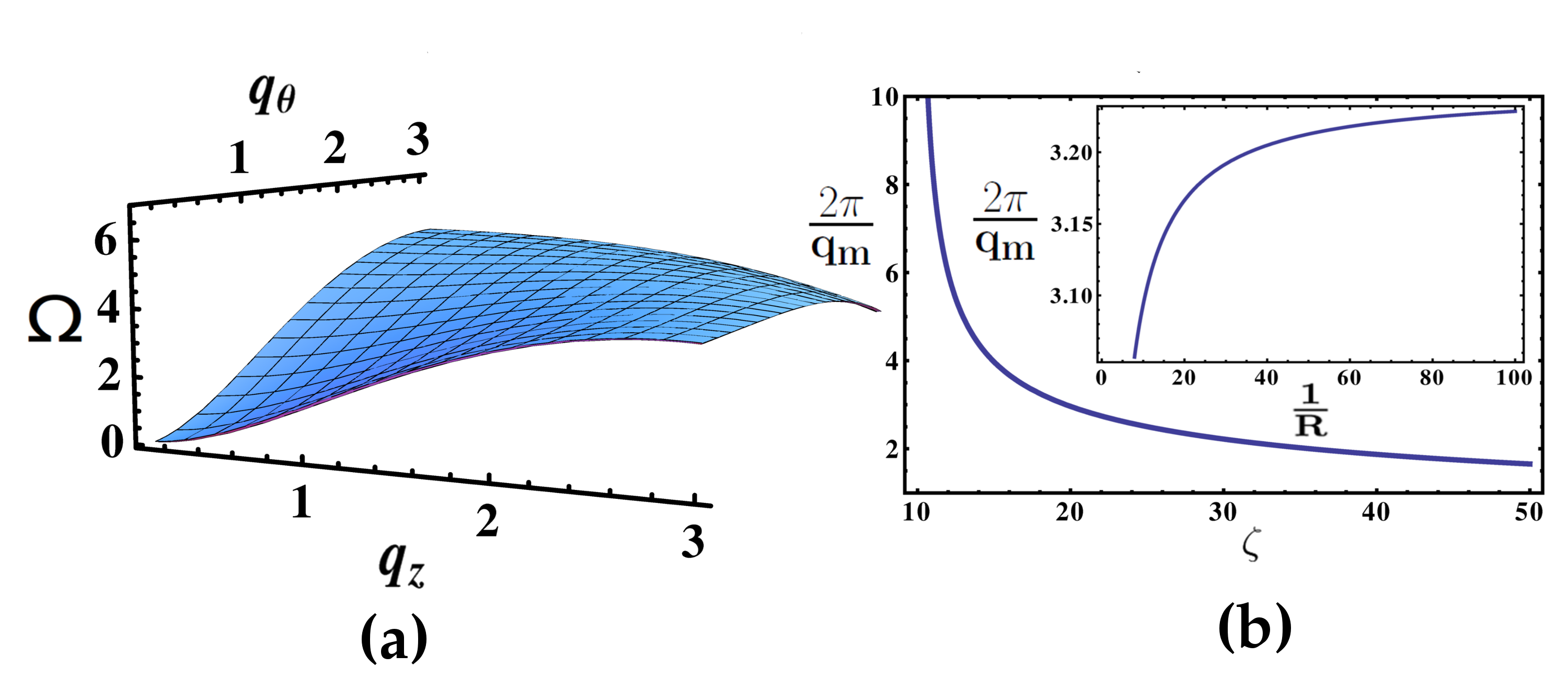}}
   \end{center}
\caption{(color online) (a) Typical growth rate $\Omega$ of the unstable mode in $\l q_\th, q_z \r$ plane. (b) Plot of  inverse of the fastest growing wavevector  $q_{m}$ with $\zeta$ and $R$ (inset).}
\label{fig1}
\end{figure*}

\begin{figure*}[ht]
  \begin{center}
    \subfigure{\includegraphics[height=3.5in]{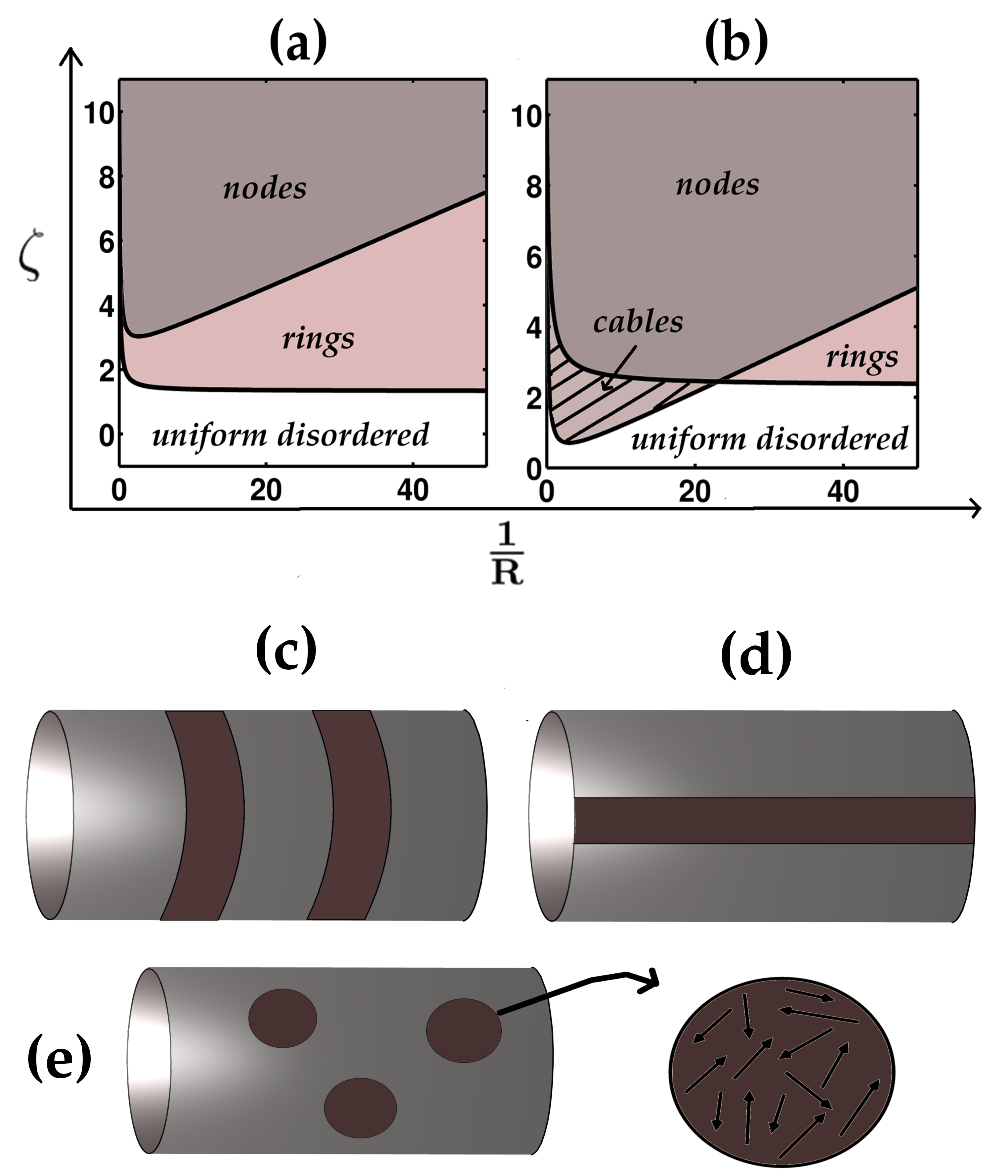}}
   \end{center}
\caption{(color online) (a)-(b) Phase diagrams in $ \zeta - R$ at low filament concentration, $\a_{i}<0$, corresponding to zero mean orientation $\langle {\bf n}\rangle =0$. The final steady state configurations obtained from a numerical solution of the dynamical equations for $c$ and ${\bf n}$, correspond to (c) rings, (d) cables and (e) nodes. (a) When $\zeta_1> \zeta_2$, the homogeneous phase is unstable to {\it rings} and {\it nodes} as $\zeta$ increases, (b) When $ \zeta_2 > \zeta_1$, the homogeneous phase is unstable to cables, at large $R$. As $R$ is lowered, the phase boundaries cross and the sequence of transitions resembles (a).}
\label{fig2}
\end{figure*}

Here we show a plot of the dispersion surface and analytic expressions for the fastest growing wave vector when mean filament density is low
$\a_i <0 $, i.e., when the homogeneous phase is orientationally disordered. As mentioned in the main text, the maxima of the dispersion surface is along $q_\th$ or $q_z$, as illustrated in Fig.\,\ref{fig1}(a),


\begin{enumerate}
\item When $\Big\vert \f{D\a_\th}{v_\th}\Big\vert = \zeta_1 > \zeta > \zeta_2 = \Big\vert \f{D\a_z}{v_z}\Big\vert$, the fastest growing wave vector ${\bf q}_m$ is along $z$, with a rate given by,

\bea
\Omega= \f{\a_z - D_{+} q_z^2}{2} + \f{1}{2} \sqrt{\l\a_z - D_{-} q_z^2\r^2 + 4 c_0 v_z \zeta q_z^2}
\label{disp}
\eea
where $D_{+}=D + K_1 + K_2$ and $D_{-}= D-K_1-K_2$. The magnitude of the most unstable wavevector is given by the maxima
of $\Omega(q_z)$ in (\ref{disp}), and is plotted as a function of $R$ and $\zeta$ in Fig.\,\ref{fig1}(b). This takes a simple form when we set $D_{-}=0$,
\be
q_m  =  \f{1}{2 D_{+}} \sqrt{\f{c_0^2 \zeta^2 v_z^2 - \a_z^2 D_{+}^2}{c_0v_z \zeta}}\,\, .
\label{qm_1}
\ee

It is instructive to look at the asymptotic forms of this equation,

\begin{enumerate}

\item As $R\rar \infty$,  $ q_m \sim  \f{1}{2 D_{+}} \sqrt{\f{c_0^2 \zeta^2 v_0^2 - \a_0^2 D_{+}^2}{c_0v_0 \zeta}}  + O(\f{1}{R})$.

\item $ q_m  \sim \sqrt{\zeta-\Big\vert \f{D \a_{z}}{v_{z}}\Big\vert}$, for $\zeta $ close to the threshold $\zeta_2=\Big\vert \f{D \a_z}{c_0 v_z} \Big\vert$. 

\item  $q_m \sim \f{c_0 v_z}{ 2 D_{+}}\sqrt{\zeta}$, for large $\zeta$. 

\end{enumerate}

The eigenvector corresponding to this instability has a large component along $c$, implying that this fastest growing instability corresponds to  concentration clumping along $\th$ direction ({\it rings}). This is verified by explicit numerical integration (see, Fig.\,\ref{fig2}(c)
for a schematic).

\item When $\Big\vert \f{D \a_z}{v_z}\Big\vert = \zeta_2 > \zeta > \zeta_1 = \Big\vert \f{D \a_\th}{v_\th}\Big\vert$, the fastest growing wavevector is along $\th$ axis. The expressions for the  dispersion relations and the magnitude of the  fastest growing wavevector in this regime is simply obtained by interchanging $z$ and $\th$ indices of the parameters in (\ref{disp}) and (\ref{qm_1}). This instability corresponds to concentration clumping along $z$ direction ({\it cables}). This is verified by explicit numerical integration (see, Fig.\,\ref{fig2}(d)
for a schematic).

\item When $\zeta > \max \l\zeta_1, \zeta_2 \r$, the dispersion surface has two maxima displayed in Fig.\,\ref{fig1}(a). A numerical solution of the dynamical equations shows that the steady state in this case consists of nodes (schematically shown in Fig.\,\ref{fig2}(e)).


\end{enumerate}

The phase diagrams in the space of $\l \zeta,R \r$ appear in Fig.\,\ref{fig2}(a)-(b) (phase boundaries computed from linear stability
analysis) and the corresponding steady state configurations are displayed in Fig.\,\ref{fig2}(c)-(e).

\subsection{High mean concentration of polar filaments}

Here we show a plot of the dispersion surface and analytic expressions for the fastest growing wave vector when mean filament density is high and uniform, i.e., either $\a_{\th}$ or $\a_z$ is positive, corresponding to orientational order along $\th$ or $z$. As mentioned in the text, without 
any loss of generality, we consider the case when $\a_{\th} $ is positive and greater than $\a_z$. In this case, the ordering is along
the $\th$-axis.

\begin{figure*}[ht]
  \begin{center}
    \subfigure{\includegraphics[height=2.15in]{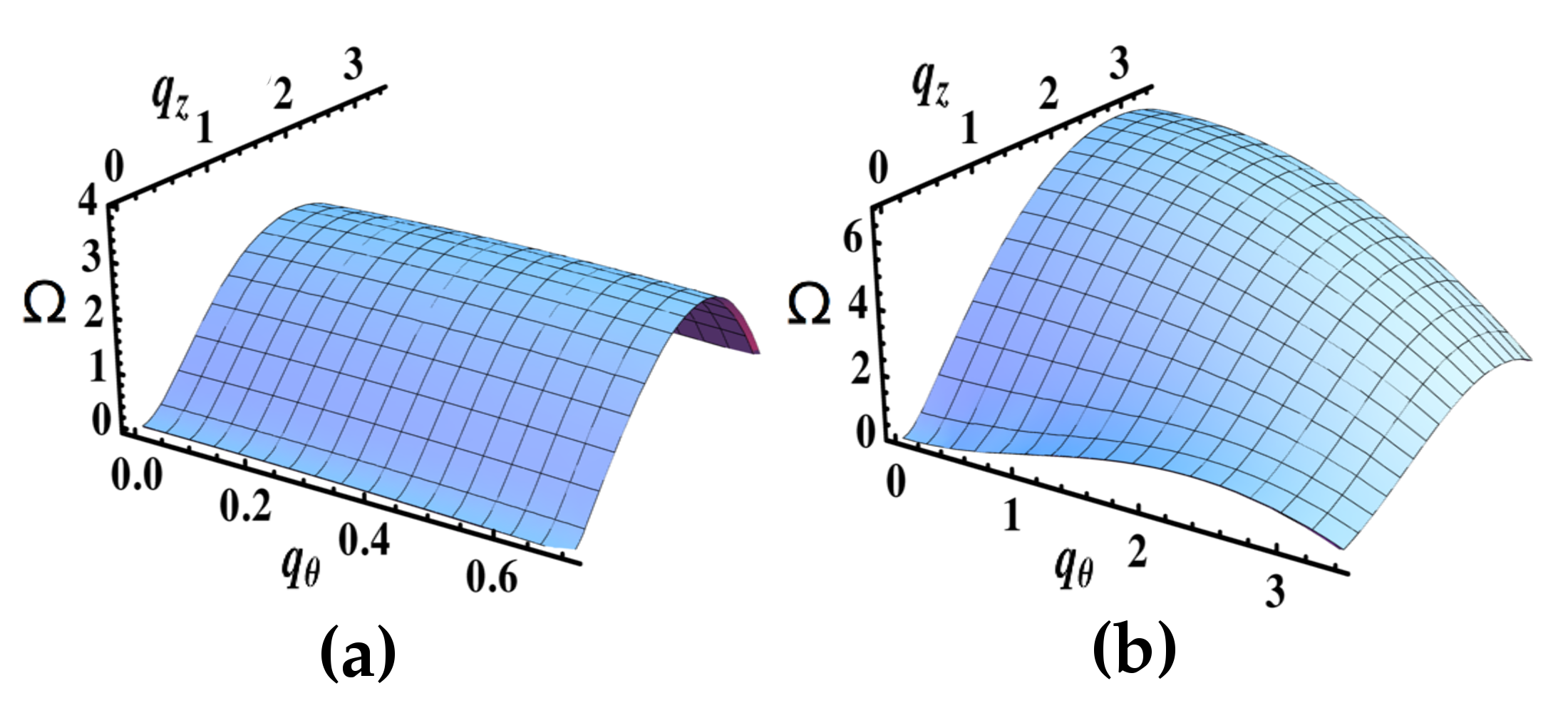}}
   \end{center}
\caption{(color online) Dispersion surface when mean density of filaments is high and the homogeneous phase is orientationally ordered along $\th$ direction. The maxima of dispersion surface is either along (a) $z$-axis or (b) both $z$ and $\th$-axes.}
\label{fig3}
\end{figure*}

As before, the maxima of the dispersion surface is along $q_\th$ or $q_z$, as illustrated in Fig.\,\ref{fig3}(a),(b),

%

\begin{enumerate}
\item When $ \f{2 \a_\th}{c_0 v_\th}=\zeta_{c2} > \zeta > \zeta_{c1}=\f{D \l \a_\th- \a_z \r}{c_0 v_z}$, the fastest growing wavevector is along $z$, with a growth rate is given by,
%

\bea
\Omega&=& \f{ \a_z-a_\th - D_{+} q_z^2 }{2} + \f{1}{2}\sqrt{\l \a_z -a_\th - D_{-} q_z^2\r^2 + 4 c_0 v_z \zeta q_z^2} \nn\\
\label{disp2}
\eea
where $D_{\pm}$ have been defined before.
%
The magnitude of the most unstable wavevector in the limit $D_{-}=0$, is given by
\be
q_m = \f{1}{2 D_{+}}\sqrt{\f{c_0^2 v_z^2 \zeta^2 - D_{+}^2 \l \a_\th-\a_z\r^2}{c_0 v_z \zeta}}
\ee
which has the following asymptotic forms,\\

 $q_m \sim \f{\sqrt{v_0 \zeta v_0}}{2 D_{+}} + O\l \f{1}{R}\r$, as $R \rar \infty$, and \\
 
 $q_m \sim \sqrt{\zeta}$ for large $\zeta$ (see Fig.\,1(c) in main text). \\

\item When $\zeta_{c1} > \zeta > \zeta_{c2}$, the most unstable wavevector is parallel to $\th$-axis. The dispersion relation  for the unstable mode and the magnitude of the fastest growing wavevector are given by, 

\bwd
\bea
\Omega= \f{-2 \a_\th +  D_{+} q_\th^2- i \sqrt{\f{\a_\th}{\b}}\l v_\th - \lambda\r}{2} + \f{1}{2} \sqrt{\l -2 \a_\th + D_{-} q_\th^2+ i \sqrt{\f{\a_\th}{\b}}\l v_\th - \lambda\r\r^2 + 4 c_0 v_\th \zeta q_\th^2}
\label{disp3}
\eea

\ewd
 
The magnitude of the  fastest growing wavevector in the limit $D_{-}=0$ is,

\be
q_m= \f{1}{2 D_{+}} \sqrt{\f{\l \f{\a_\th}{\b}(v_\th -\lambda)^2- 4 c_0 \zeta v_\th \r^2 - 16 D_{+}^2 \a_\th^2}{4 c_0 v_\th \zeta -\f{\a_\th}{\b}(v_\th -\lambda)^2 }}
\ee

\end{enumerate}

The final steady states corresponding to (a)-(b) are obtained from a numerical solution of the dynamical equations and correspond to stationary rings and moving cables. When $\zeta > \max(\zeta_{c1},\zeta_{c2})$, the numerical solution shows that 
asters/nodes are the final steady state configurations. The phase diagrams appear in Fig.\,1(a)-(b) in the main text.
\\

\section{Time sequences of configurations from numerical integration}

\begin{figure*}[ht]
  \begin{center}
    \subfigure{\includegraphics[height=2.15in]{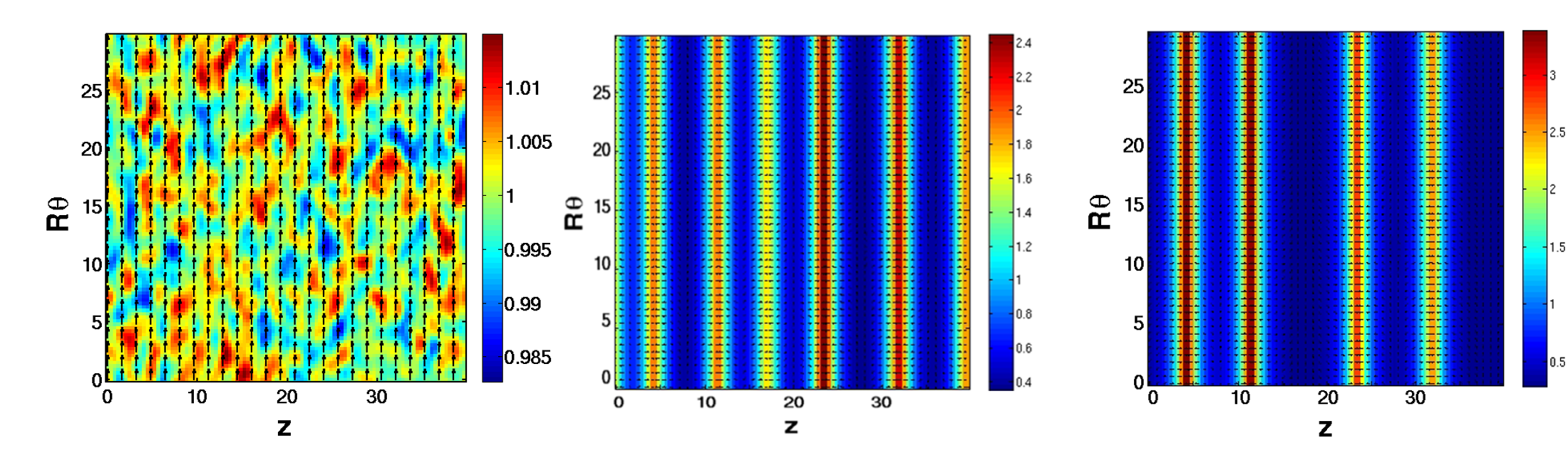}}
   \end{center}
\caption{(color online) Time evolution of configurations starting from an (a) initial homogeneous, orientationally ordered configuration
along $\th$, which shows an (b) early time spinodal instability toward ring formation, followed by a (c) slow coarsening regime. Colorbar denotes the concentration of actin filaments, in units of mean concentration $c_0$.}
\label{fig4}
\end{figure*}

In Fig. \ref{fig4} we show snapshots from the numerical integration of the dynamical equations when filament density is high, such that $\a_\th > \max\l \a_z,0\r$ and $\zeta_{c2} > \zeta > \zeta_{c1}$.  The initial homogeneous, orientationally ordered configuration
along $\th$, is immediately unstable to the formation of rings. This spinodal instability happens exponentially fast. The subsequent
dynamics is a slow coarsening of the rings, which involves ring diffusion and coalescence, resulting in a reduction in the number of rings
over time.


\newpage

\newpage

\newpage

\newpage

